\documentclass[aps,preprint,amsfonts,amssymb,amsmath]{revtex4}

\newcommand{\be}{\begin{eqnarray}}
\newcommand{\ee}{\end{eqnarray}}
\newcommand{\bm}{\boldsymbol}
\newcommand{\balpha}{\mbox{\boldmath $\alpha$}}

\newcommand{\bsigma}{\mbox{\boldmath $\sigma$}}

\begin{document}

\title{Dirac(-Pauli), Fokker-Planck equations and exceptional Laguerre polynomials}

\author{Choon-Lin Ho}
\affiliation{Department of Physics, Tamkang University,
 Tamsui 251, Taiwan, R.O.C.}


\begin{abstract}

An interesting discovery in the last two years in the field of
mathematical physics has been the exceptional $X_\ell$ Laguerre
and Jacobi polynomials.  Unlike the well-known classical
orthogonal polynomials which start with constant terms, these new
polynomials have lowest degree $\ell=1,2,\ldots$, and yet they
form complete set with respect to some positive-definite measure.
While the mathematical properties of these new $X_\ell$
polynomials deserve further analysis, it is also of interest to
see if they play any role in physical systems.  In this paper we
indicate some physical models in which these new polynomials
appear as the main part of the eigenfunctions. The systems we
consider include the Dirac equations coupled minimally and
non-minimally with some external fields, and the Fokker-Planck
equations.  The systems presented here have enlarged the number of
exactly solvable physical systems known so far.

\end{abstract}




 \maketitle

\section{Introduction}

The past two years have witnessed some interesting developments in
the area of exactly solvable models in quantum mechanics: the
number of exactly solvable shape-invariant models has been greatly
increased owing to the discovery of new types of orthogonal
polynomials, called the exceptional $X_\ell$ polynomials.  Two
families of such polynomials, namely, the Laguerre- and
Jacobi-type $X_1$ polynomials, corresponding to $\ell=1$, were
first proposed by G\'omez-Ullate et al. in \cite{GKM1}, within the
Sturm-Liouville theory, as solutions of second-order eigenvalue
equations with rational coefficients. Unlike the classical
orthogonal polynomials, these new polynomials have the remarkable
properties that they still form complete set with respect to some
positive-definite measure, although they start with a linear
polynomials instead of a constant. The results in \cite{GKM1} were
then reformulated by Quesne in the framework of quantum mechanics
and shape-invariant potentials, first in \cite{Que1} by the point
canonical transformation method, and then in \cite{Que2} by
supersymmetric (SUSY) method \cite{SUSY} (or the Darboux-Crum
transformation \cite{DC}). Soon after these works, such kind of
exceptional polynomials were generalized by Odake and Sasaki to
all integral $\ell=1,2,\ldots$ \cite{OS1} (the case of $\ell=2$
was also discussed in \cite{Que2}). By construction these new
polynomials satisfy the Schr\"odinger equation and yet they start
at degree $\ell>0$ instead of the degree zero constant term. Thus
they are not constrained by Bochner's theorem \cite{Bochner},
which states that the orthogonal polynomials (starting with degree
0) satisfying a second order differential equations can only be
the classical orthogonal polynomials, i.e., the Hermite, Laguerre,
Jacobi and Bessel polynomials.

Later, equivalent but much simpler looking forms of the Laguerre-
and Jacobi-type $X_{\ell}$ polynomials than those originally
presented in \cite{OS1} were given in \cite{HOS}.  These nice
forms were derived based on an analysis of the second order
differential equations for the $X_{\ell}$ polynomials within the
framework of the Fuchsian differential equations in the entire
complex $x$-plane.  They allow us to study in-depth some important
properties of the $X_\ell$ polynomials, such as the actions of the
forward and backward shift operators on the $X_{\ell}$
polynomials, Gram-Schmidt orthonormalization for the algebraic
construction of the $X_{\ell}$ polynomials, Rodrigues formulas,
and the generating functions of these new polynomials.

Recently, in \cite{GKM2} the $X_1$ Laguerre polynomials in
\cite{GKM1} are generalized to the $X_\ell$ Laguerre polynomials
with higher $\ell$ based on the Darboux-Crum transformation. Then
in \cite{STZ} such transformation was successfully employed to
generate the $X_\ell$ Jabobi as well as the $X_\ell$ Laguerre
polynomials as given in \cite{HOS}.

While the mathematical properties of these new $X_\ell$
polynomials deserve further analysis, it is also of interest to
see if they play any role in physical systems. It is the purpose
of the present work to indicate some physical models that involve
these new polynomials.  We shall be mainly concerned with the
$X_\ell$ Laguerre polynomials for clarity of presentation.  Models
that are linked with the $X_\ell$ Jacobi polynomials will be
briefly mentioned at the end.

The plan of the paper is as follows. First we review the deformed
radial oscillators associated with the exceptional $X_\ell$
Laguerre polynomials in Sect.~II.  In Sect.~III the Dirac equation
minimally coupled with an external magnetic field is considered.
We present the forms of the vector potentials such that the
eigenfunctions of the Dirac equation are related to the $X_\ell$
Laguerre polynomials. Dirac equations with non-minimal couplings
are then mentioned in Sect.~IV, and the Fokker-Planck equations
are considered in Sect.~V.  Sect.~VI concludes the paper.

\section{Exceptional $X_\ell$ Laguerre polynomials}

Consider a generic one-dimensional quantum mechanical system
described by a Hamiltonian $H=-d^2/dx^2 +V_0(x)$. Suppose the
ground state is given by the wave function $\phi_0(x)$ with zero
energy: $H\phi_0=0$. By the well known oscillation theorem
$\phi_0$ is nodeless, and thus can be written as $\phi_0\equiv
e^{W_0(x)}$, where $W_0(x)$ is a regular function of $x$. This
implies that the function $W_0(x)$ completely determines the
potential $V_0$ : $V_0={W_0^\prime}^2 + W_0^{\prime\prime}$ (the
prime here denotes derivatives with respect to $x$). Thus $W_0(x)$
is sometimes called a prepotential. Consequently, the Hamiltonian
can be factorized as $H\equiv H_0^{(+)}=A_0^{-}A_0^+$, with
$A_0^\pm\equiv \pm d/dx  -W_0^\prime$. It is trivial to verify
$A_0^+\phi_0(x)=0$.

It is a remarkable fact that most well-known one-dimensional
exactly solvable systems possess a property called shape
invariance \cite{SI}. This means the Hamiltonian $H^{(-)}_0$,
defined by $H_0^{(-)}=A_0^{+}A_0^-$ with potential ${W_0^\prime}^2
- W_0^{\prime\prime}$, is related to $H_0^{(+)}$ by the relation $
H_0^{(-)}(\bm{\lambda})
 =H_0^{(+)}(\bm{\lambda}+\bm{\delta})
  +\mathcal{E}_1(\bm{\lambda})$.
Here $\bm{\lambda}=(\lambda_1,\lambda_2,\ldots)$ is a set of
parameters of the Hamiltonian $H=H(\bm{\lambda})$, $\bm{\delta}$
is a certain shift of these parameters, and
$\mathcal{E}_1(\bm{\lambda})$ is some function of $\bm{\lambda}$.
The mapping that relates the potentials $H_0^{(+)}$ and
$H_0^{(-)}$ is called the Darboux-Crum transformation \cite{DC}
($H_0^{(+)}$ and $H_0^{(-)}$ are also called SUSY partners in the
context of SUSY quantum mechanics \cite{SUSY}).  Shape invariance
is a sufficient condition \cite{Ho1} that enables one to determine
the eigenvalues and the corresponding eigenfunctions of $
H_0^{(+)}(\bm{\lambda})$ exactly (see \cite{SUSY} for details).
Specifically, we have
\begin{eqnarray}
\mathcal{E}_0(\bm{\lambda})&=&0,~~\mathcal{E}_m(\bm{\lambda})
  =\sum_{k=0}^{n-1}\mathcal{E}_1(\bm{\lambda}+k\bm{\delta}),\\
~~\phi_m(x;\bm{\lambda})&\propto &
  A_0^-(\bm{\lambda})
  A_0^-(\bm{\lambda}+\bm{\delta})\cdots
  A_0^-(\bm{\lambda}+(m-1)\bm{\delta})
  \times e^{W_0(x;\bm{\lambda}+m\bm{\delta})},~~ m=1,2,\ldots.
  \end{eqnarray}

The new shape-invariant systems related to the exceptional
polynomials are determined by certain prepotentials $W_{\ell}$
($\ell=1,2,\ldots$), which are obtained by deforming some shape
invariant prepotentials $W_0$ \cite{OS1}. The $\ell=0$ case
corresponds to the original system. Three families of such exactly
solvable deformed systems were presented in \cite{OS1}, which
correspond to deforming the radial oscillator and the
trigonometric/hyperbolic Darboux-P\"oschl-Teller potentials in
terms of their respective eigenfunctions.

For the purpose of this paper, we shall only consider the
exceptional $X_{\ell}$ Laguerre polynomials, which appear in the
deformed radial oscillator potentials.  The original radial
oscillator potential is generated by the prepotential
 \be
W_0(x;g)=-\frac{\omega x^2}{2}+g\log x,~~0<x<\infty.
 \label{w0}
 \ee
Here $\bm{\lambda} =g>0$ and $\omega>0$. The Hamiltonian is
 \be
 H_0^{(+)}(g)=A^-(g)A^+(g)
 =-\frac{d^2}{dx^2}+\omega^2
 x^2+\frac{g(g-1)}{x^2}-(2g+1)\omega.\label{H0-osc}
\ee This potential is shape-invariant with shift parameter
$\bm{\delta}=1$ and $\mathcal{E}_1(g)=4\omega$: $
H_0^{(-)}(g)=H_0^{(+)}(g+1)+4\omega$. The eigen-energies and
eigenfunctions are ($n=0,1,2,\ldots$)
 \be
  \mathcal{E}_n(g)=4n \omega;~~~\phi_n(x;g)=e^{W_0(x;g)} P_n
  (\eta;g),~~P_n(\eta;g)\equiv
  L_n^{(g-\frac{1}{2})}(\eta),
  \ee
where $\eta(x)\equiv \omega x^2$ is one of the so-called
sinusoidal coordinates \cite{OS2}.

Below we shall present radial oscillators related to the $X_\ell$
Laguerre polynomials. We first treat them as systems deformed from
the original radial oscillator by appropriate deforming functions.
Then we show how they can also be considered as the Darboux-Crum
(or SUSY) partner of the original radial oscillator.

\subsection{Deformed radial oscillators}

The Hamiltonian $H_\ell^{(+)}(g)$ of the deformed radial
oscillator is
 \be
H^{(+)}_{\ell}(g)=-\frac{d^2}{dx^2} +W_\ell^{\prime 2} (x;g) +
W_\ell^{\prime\prime}(x;g),\label{H-SO}
 \ee
where $W_\ell$ is given by
\begin{equation}
W_\ell(x;g)= -\frac{\omega x^2}{2}+(g+\ell)\log x
+\log\frac{\xi_\ell(\eta;g+1)}{\xi_\ell(\eta;g)}.\label{wl-osc}
\end{equation}
Here  $\xi_\ell(\eta;g)$ is a deforming function. It turns out
there are two possible sets of deforming functions
$\xi_\ell(\eta;g)$, thus giving rise to two sets of infinitely
many exceptional Laguerre polynomials, termed L1 and L2 type
\cite{OS1,HOS}. These $\xi_\ell$ are given by
 \be
  \xi_{\ell}(\eta;g)=
  \left\{
  \begin{array}{ll}
  L_{\ell}^{(g+\ell-\frac32)}(-\eta)&:\text{L1}\\
  L_{\ell}^{(-g-\ell-\frac12)}(\eta)&:\text{L2}.
  \end{array}\right.
  \label{xiL}
\ee

For both types of $\xi_\ell$, the eigen-energies are
$\mathcal{E}_{\ell,n}(g)=\mathcal{E}_n(g+\ell)=4n\omega$, which
are independent of $g$ and $\ell$. Hence the deformed radial
oscillator is iso-spectral to the ordinary radial oscillator. The
eigenfunctions are given by
 \be
\phi_{\ell,n}(x;g)=\frac{e^{-\frac{1}{2}\omega
x^2}x^{g+\ell}}{\xi_{\ell}(\eta;g)} P_{\ell,n}(\eta;g),
 \label{phi-l}
 \ee
where the corresponding exceptional Laguerre polynomials
$P_{\ell,n}(\eta;g)$ ($\ell=1,2,\ldots$, $n=0,1,2,\ldots$) can be
expressed as a bilinear form of the original Laguerre polynomials
and the deforming polynomials, as given in \cite{HOS}:
\begin{equation}
  P_{\ell,n}(\eta;g)=
  \left\{
  \begin{array}{ll}
  \xi_{\ell}(\eta;g+1)P_n(\eta;g+\ell-1)
  -\xi_{\ell}(\eta;g)\partial_{\eta}
  P_n(\eta;g+\ell-1)&:\text{L1}\\[2pt]
  (n+g+\frac12)^{-1}\bigl((g+\frac12)
  \xi_{\ell}(\eta;g+1)P_n(\eta;g+\ell+1)\\
  \phantom{(n+g+\frac12)^{-1}\bigl(}\ \quad
  +\eta\xi_{\ell}(\eta;g)\partial_{\eta}P_n(\eta;g+\ell+1)
  \bigr)&:\text{L2}.
  \end{array}\right.
  \label{XL}
\end{equation}
 The $X_{\ell}$ polynomials
$P_{\ell,n}(\eta;g)$ are degree $\ell+n$ polynomials in $\eta$ and
start at degree $\ell$: $P_{\ell,0}(\eta;g)=\xi_{\ell}(\eta;g+1)$.
They are orthogonal with respect to certain weight functions,
which are deformations of the weight function for the Laguerre
polynomials (for details, see \cite{HOS}).

\subsection{Darboux-Crum pairs}

As mentioned in Sect.~1, the new exceptional orthogonal
polynomials have recently been re-derived from the corresponding
ordinary polynomials based on the Darboux-Crum transformation
\cite{GKM2,STZ}.  Below we shall reconsider the case of the
deformed radial oscillator, following the approach of \cite{STZ}.

Instead of finding the prepotential $W_\ell$ that gives
$H_0^{(+)}$ as the deformed oscillator, as was done in the
previous subsection, one determines a new prepotential $W_\ell$ so
that it is $H_0^{(-)}$ that gives the deformed oscillator while
$H_0^{(+)}$ is related to the ordinary oscillator.

The new prepotential $W_\ell(x;g)$ are \cite{STZ}
\begin{eqnarray}
W_\ell(x;g)&= & \left\{
  \begin{array}{ll}
\frac{1}{2}\omega x^2+(g+\ell-1)\log x
+\log\xi_\ell(\eta(x);g),\quad g>1/2, ~&: \text{L1}\\
 -\frac{1}{2}\omega x^2-(g+\ell)\log x
+\log\xi_\ell(\eta(x);g),\quad g>-1/2,~&: \text{L2}
 \end{array}\right.
 \label{wl-DC}
\end{eqnarray}
where $\xi_\ell(\eta;g)$ are as given in (\ref{xiL}). Now consider
the pairs of Hamiltonians $\mathcal{H}^{(+)}_\ell(g)$ and
$\mathcal{H}^{(-)}_\ell(g)$ ($\ell=1,2,\ldots$) defined by
 \be
\mathcal{H}^{(+)}_\ell(g) &=&
\mathcal{A}^-_\ell(g)\mathcal{A}^+_\ell(g),~~~
\mathcal{H}^{(-)}_\ell(g)= \mathcal{A}^+_\ell(g) \mathcal{A}^-_\ell(g),\label{Hl-DC}\\
\mathcal{A}^\pm_\ell(g) &=& \pm\frac{d}{dx}-W^\prime_\ell(x;g).
 \ee
 Using the
differential equation for the Laguerre polynomial, the
Hamiltonians $\mathcal{H}^{(+)}_\ell(g)$ can be shown to be
related to that of the radial oscillator $H_0^{(+)}(g)$ in
(\ref{H0-osc}) as
\begin{align}
\mathcal{H}^{(+)}_\ell(g)= \left\{
  \begin{array}{ll}
H_0^{(+)}(g+\ell-1)+2(2g+4\ell-1)\,\omega~~&: \text{L1}\\
H_0^{(+)}(g+\ell+1)+2(2g+1)\,\omega~~&: \text{L2} .
  \end{array}\right.
  \label{Hl-DC-1}
\end{align}
The partner Hamiltonians are found to equal to the Sasaki-Odake
Hamiltonian $H^{(+)}_{\ell}(g)$ in (\ref{H-SO}) up to additive
constants:
\begin{align}
\mathcal{H}^{(-)}_\ell(g)= \left\{
  \begin{array}{ll}
H_\ell^{(+)}(g)+2(2g+4\ell-1)\,\omega~~&: \text{L1}\\
H_\ell^{(+)}(g)+2(2g+1)\,\omega~~&: \text{L2} .
  \end{array}\right.
\end{align}

It is shown in \cite{STZ} that the partner Hamiltonian
$\mathcal{H}^{(-)}_\ell(g)$ are exactly iso-spectral to the radial
oscillator Hamiltonian $\mathcal{H}^{(+)}_\ell(g)$, which have the
following eigenvalues and the corresponding eigenfunctions
($\eta=\omega x^2$):
\begin{alignat}{2}
&\text{L1:}\quad
\mathcal{E}_{\ell,n}^{(+)}(g)=4\left(n+g+2\ell-\frac12\right)\,\omega,
\quad &\phi_{\ell,n}^{(+)}(x;g) =e^{-\frac{1}{2}\omega
x^2}x^{g+\ell-1}L_n^{(g+\ell-\frac32)}(\eta),
\label{eigen+1}\\
&\text{L2:}\quad
\mathcal{E}_{\ell,g}^{(+)}(g)=4\left(n+g+\frac12\right)\,\omega,\quad
&\phi_{\ell,n}^{(+)}(x;g) =e^{-\frac{1}{2}\omega
x^2}x^{g+\ell+1}L_n^{(g+\ell+\frac12)}(\eta). \label{eigen+2}
\end{alignat}
The eigenfunctions $\phi_{\ell,n}^{(-)}(x;g)$ of
$\mathcal{H}^{(-)}_\ell(g)$ are obtained by applying the operator
$\mathcal{A}^+_\ell(g)$ on $\phi_{\ell,n}^{(+)}(x;g)$:
$\phi_{\ell,n}^{(-)}(x;g)= \mathcal{A}^+_\ell(g)
\phi_{\ell,n}^{(+)}(x;g)$.  It turns out that
$\phi_{\ell,n}^{(-)}(x;g)$, up to multiplicative constants, are
just the eigenfunctions $\phi_{\ell,n}(\eta;g)$ in (\ref{phi-l})
and (\ref{XL}) of the Sasaki-Odake Hamiltonian.

\section{Dirac equation with magnetic field in cylindrical coordinates}

Having reviewed the deformed radial oscillators associated with
the exceptional $X_\ell$ Laguerre polynomials, we will like to see
if there are physical systems in which these new polynomials could
play a role.  It turns out that there are indeed such systems.  In
what follows we shall indicate some of them.

First let us consider the Dirac equation in 2+1 dimensions
coupling minimally with a cylindrically symmetric magnetic field.
The discussion can be extended to 3+1 dimensions where the
magnetic field does not depend on the variable $z$.  The Dirac
equation for a unit positive ($q=1$) charged particle minimally
coupled to an magnetic field $\bf{B}=\nabla \times \bf{A}$ has the
form (we set $c=\hbar=1$)
\begin{eqnarray}
H_D \Psi({\bf r})=E\Psi({\bf r})\qquad\nonumber\\
  H_D = {\bsigma}\cdot{\bf (p-A)} + \sigma_3 M,\label{H-D}
\end{eqnarray}
where $H_D$ is the Dirac Hamiltonian, $\bm{p}= -i\nabla$ is the
momentum operator, $E$ and $M$ are energy and rest mass of the
particle, and $\bsigma$ are the Pauli matrices.

We shall consider magnetic field which is cylindrically symmetric.
Then the vector potential has only $\phi$-component:
\begin{equation}
{\bm{A}}({\bf r})=A_\phi(r)\hat{\phi},\ \ r=|\bm{r}|.
\end{equation}
Instead of the variable $x$, we shall use the conventional
notation $r$ for the radial variable here, and in Sect.~IV.A and
B.  The magnetic field is
 \be
  B_3(r)=\frac{1}{r}\frac{d}{dr}\left(rA_\phi(r)\right).
 \ee
  The wave function is
taken to have the form
\begin{eqnarray}
\psi_m(r, \phi) = \left( \begin{array}{c}
f_+(r)e^{im\phi}\\
-if_-(r)e^{i(m+1)\phi}
\end{array}\right)
\label{wf1}
\end{eqnarray}
with integral number $m$. The function $\psi_m(r,\phi)$ is an
eigenfunction of the conserved total angular momentum $J_3=L_3 +
S_3 = -i\partial/\partial\phi + \sigma_3/2$ with eigenvalue
$j=m+1/2$.  It should be reminded that $m$ is not a good quantum
number. This is evident from the fact that the two components of
$\psi_m$ depend on the integer $m$ in an asymmetric way. Only the
eigenvalues $j$ of the conserved total angular momentum $J_3$ are
physically meaningful.

From the identities
 \be
 {\bsigma}\cdot {\bf p}&=&i({\bsigma}\cdot {\hat {\bf
r}})\left(-\partial_r+ \frac {1}{r}({\bsigma}\cdot {\bf
L})\right),\label{id}\\
 \bsigma\cdot {\bf A} &=&i(\bsigma\cdot {\hat {\bf
r}})\sigma_3 A_\phi, \label{id-1}
 \ee
 one gets
 \be
{\bsigma}\cdot {\bf (p- A)})=i({\bsigma}\cdot {\hat {\bf
r}})\left(-\partial_r+ \frac{1}{r}\left(\sigma_3
J_3-\frac{1}{2}\right)- \sigma_3 A_\phi\right).\label{p-A}
 \ee
Upon using the relation
 \be
\left(\bsigma\cdot {\hat {\bf r}}\right)\left( \begin{array}{c}
F e^{im\phi}\\
G e^{i(m+1)\phi}
\end{array}\right)=\left( \begin{array}{c}
G e^{im\phi}\\
F e^{i(m+1)\phi}
\end{array}\right),
\ee
 one can reduce (\ref{H-D}) to
 \be
 \left(\frac{d}{dr} -
\frac{m+\frac{1}{2}}{r}
+ A_\phi\right)f_+ &=& \left(E + M\right)f_-,\label{f+}\\
\left(-\frac{d}{dr} - \frac{m+\frac{1}{2}}{r} + A_\phi\right)f_-
&=& \left(E - M\right)f_+. \label{f-}
 \ee
This shows that $f_+$ and $f_-$ forms a one-dimensional SUSY
pairs.  If we take the prepotential $W$ such that
 \be
W^\prime=\frac{g}{r} - A_\phi, ~~~g\equiv m+\frac{1}{2},\label{w1}
 \ee
 and $A^\pm=\pm d/dr - W^\prime$ (as usual, the
prime here denotes derivatives with respect to the basic variable,
which is $r$ in this case),  then eqs.(\ref{f+}) and (\ref{f-})
become
 \be
A^-A^+f_+&=&\left(E^2 -M^2\right)f_+,\label{SUSY+}\\
A^+A^-f_-&=&\left(E^2 -M^2\right)f_-,\label{SUSY-}
 \ee
Explicitly, the above equations read
 \be
\left(-\frac{d^2}{dr^2} + W^{\prime 2} \pm
W^{\prime\prime}\right)f_\pm = \left(E^2 - M^2\right)f_\pm.
\label{susy-1}
 \ee
The ground state, with $E^2=M^2$,  is given by one of the
following two sets of equations:
 \be
A^+ f_+^{(0)}(r)&=& 0,~~~ f_-^{(0)}(r)=0,\\
 A^- f_-^{(0)}(r)&=&
0, ~~~f_+^{(0)}(r)=0,
 \ee
depending on which solution is normalizable. The solutions are
generally given by
 \be
f^{(0)}_\pm\propto r^{\pm (m+\frac{1}{2})} \exp\left(\mp\int dr
A_\phi\right).
 \ee
 To be specific, we consider
the situation where $m\geq 0$ and $\int dr A_\phi >0$, so that
$f_+^{(0)}$ is normalizable, and $f_-^{(0)}=0$. The other
situation can be discussed similarly.

Eq.~(\ref{w1}) relates $A_\phi(r)$ and $W(r)$. This gives a way to
obtain $A_\phi$ that defines exactly solvable model. Particularly,
from the Table~(4.1) in \cite{SUSY}, one concludes that there are
three forms of $A_\phi$ giving exact solutions of the problem
based on the conventional classical orthogonal polynomials:

~~~~~~~~~~i)   oscillator-like :  $A_\phi(r)\propto r $~;

~~~~~~~~~~ii)  Coulomb potential-like :   $A_\phi(r)\propto {\rm
constant} $~;

~~~~~~~~~~iii) zero field-like :  $A_\phi(r)\propto 1/r $~.

\noindent Case (i) corresponds simply to the well-known Landau
level problem.

Now with the discovery of the exceptional Laguerre polynomials,
one can find an infinite family of vector potentials $A_\phi$ that
give the oscillator-like spectra. These are the deformed Landau
systems.  There are, however, two different types of deformed
Landau systems, depending on whether we choose $A^-A^+$ in
(\ref{SUSY+}) to correspond to $H_\ell^{(+)}(g)$ in (\ref{H-SO}),
or to $\mathcal{H}_\ell^{(+)}(g)$ in (\ref{Hl-DC}), or
equivalently, (\ref{Hl-DC-1}).

If we choose  $A^-A^+$ to correspond to $H_\ell^{(+)}(g)$, then
from (\ref{w1}) and (\ref{wl-osc}) we get the required vector
potential (we add a superscript to indicate its family $\ell$)
 \be
A^{(\ell)}_\phi (r) &=&\frac{m+\frac{1}{2}}{r} -
W_\ell^\prime,\nonumber\\
&=& \omega r - \left[\frac{\ell}{r}
+\frac{\xi^\prime_{\ell}(\eta;g+1)}{\xi_{\ell}(\eta;g+1)}
-\frac{\xi^\prime_{\ell}(\eta;g)}{\xi_{\ell}(\eta;g)}\right].
\label{A-ell}
 \ee
As before, the $\xi_\ell$ is given by (\ref{xiL}) for the L1 and
the L2 case.  These deformed systems are iso-spectral to the
relativistic Landau system in case (i) mentioned before: the
eigen-energies being
\begin{equation}
\mathcal{E}_{\ell,n}(g)= E^2_{\ell,n}-M^2=4n\omega,
\end{equation}
which is independent of $g$ and $\ell$.  The eigenfunction $f_+$
is given by $\phi_{\ell,n}(r;g)$ in (\ref{phi-l}): $f_+\propto
\phi_{\ell,n}(r;g)$, and $f_-$ is obtained from $f_+$ by
(\ref{f+}): $f_-\propto A^+f_+$. Thus both the upper and lower
components of the eigenfunctions are related to the exceptional
orthogonal polynomials in this case.

In the other choice of the vector potential, i.e., with $A^-A^+$
corresponding to $\mathcal{H}_\ell^{(+)}(g)$, only the lower
component $f_-$ involves the new polynomials.  This time
$A^{(\ell)}_\phi(r)$ is determined by (\ref{w1}) and
(\ref{wl-DC}):
\begin{align}
A^{(\ell)}_\phi(r)= \left\{
  \begin{array}{ll}
-\omega r - \frac{\ell-1}{r}
-\frac{\xi^\prime_{\ell}(\eta;g)}{\xi_{\ell}(\eta;g)}~~&: \text{L1}\\
\omega r + \frac{2g + \ell}{r}
-\frac{\xi^\prime_{\ell}(\eta;g)}{\xi_{\ell}(\eta;g)}~~&:
\text{L2} .
  \end{array}\right.
\end{align}
The eigen-energies are given by $\mathcal{E}_{\ell,n}^{(+)}(g)$ in
(\ref{eigen+1}) and (\ref{eigen+2}),
\begin{align}
\mathcal{E}_{\ell,n}^{(+)}(g)= E^2_{\ell,n}-M^2= \left\{
  \begin{array}{ll}
4\left(n+g+2\ell-\frac12\right)\,\omega~~&: \text{L1}\\
4\left(n+g+\frac12\right)\,\omega~~&: \text{L2} .
  \end{array}\right.
\end{align}
The eigenfunction $f_+$ is given by $\phi^{(+)}_{\ell,n}(r;g)$ in
(\ref{eigen+1}) and (\ref{eigen+2}): $f_+\propto
\phi^{(+)}_{\ell,n}(x;g)$, and $f_-$ is again obtained from $f_+$
by (\ref{f+}): $f_-\propto \phi^{(-)}_{\ell,n}=\mathcal{A}^+_\ell
f_+$. Hence, $f_-$ is related to the new exceptional orthogonal
polynomial, as discussed in the statements below (\ref{eigen+2}).
Note here that in this case the ground state energy is not zero,
and hence $f_-$ need not vanish.  Such situation is called broken
supersymmetry in SUSY quantum mechanics \cite{SUSY}.

The above discussion can be straightforwardly extended to the case
of two-dimensional Pauli equation. This is so as the square of the
Dirac equation is related to the Pauli equation. Specifically, we
have $H_D^2=H_P+M^2$, where the Pauli Hamiltonian $H_P$ is
 \be
H_P\equiv ({\bf p-A})^2 - \sigma_3 (\nabla\times \bm{A})_3.
 \ee

\section{Dirac equations with non-minimal coupling}

The example discussed in the previous section illustrates how
physical Dirac systems whose eigenfunctions are related to the
exceptional orthogonal polynomials can be constructed.  The
construction relies mainly on the fact that the two radial
components of the wave function form a SUSY pair as in (\ref{f+})
and (\ref{f-}).  Thus, as long as a Dirac equation can be reduced
to such a form, this construction applies and one can obtain new
Dirac systems that involve the new polynomials.  In this section
we indicate three more Dirac-type equations which allow such
construction. Unlike the the model discussed in Sect.~III, these
three systems involve non-minimal couplings between the fermion
and the external electromagnetic fields.

\subsection{Dirac-Pauli equation with electric field in spherical
coordinates}

Consider a neutral fermion interacting with electromagnetic fields
through its anomalous magnetic moment. The relativistic wave
equation that describe such interaction is called the Dirac-Pauli
equation \cite{D-P,Ho2}. As shown in \cite{Ho2}, this equation
includes as a special case the so-called Dirac oscillator, which
has attracted much attention in recent years \cite{Dirac-osc}.

In this section we shall consider the case where the external
field is purely electrical.  The Dirac-Pauli equation that
describes the motion of a neutral fermion of spin-1/2 with mass
$M$ and an anomalous magnetic moment $\mu$ in an external electric
field $\bf E$ is given by $H_{DP} \Psi=E \Psi$, with the
Hamiltonian
 \be
H_{DP}={\balpha}\cdot{\bf p}+i\mu \beta{\balpha}\cdot{\bf E}
+\beta M.\label{H}
 \ee
Here $\bsigma$ and $\beta$ are the Dirac matrices which in the
standard representation are given by
 \be
 {\balpha} =
\left( \begin{array}{cc} 0 & {\bsigma}\\
{\bsigma} & 0
\end{array}\right),~~~~~
\beta= \left( \begin{array}{cc} 1 & 0\\ 0 & -1
\end{array}\right),
\ee where $\bsigma$ are the Pauli matrices. We also define
$\Psi=(\chi, \varphi)^t$, where $t$ denotes transpose, and both
$\chi$ and $\varphi$  are two-component spinors. Then the
Dirac--Pauli equation becomes \be
{\bsigma}\cdot({\bf p}-i\mu {\bf E})\chi &=&(E+ M)\varphi ,\nonumber\\
 {\bsigma}\cdot({\bf p}+i\mu {\bf
E})\varphi &=&(E-M)\chi .\label{H1}
 \ee
This exhibits the intrinsic SUSY structure of the system.  As in
the previous case, this allows us to construct exactly solvable
Dirac-Pauli equations which are related to the exceptional
polynomials.  This we shall do below in the spherical and
cylindrical coordinates.

First let us consider central electric field ${\bf E}=E_r (r)
{\hat{\bf r}}$. In this case, one can choose a complete set of
observables to be $\{H,{\bf J}^2,J_z,{\bf S}^2=3/4,K\}$. Here $\bf
J$ is the total angular momentum ${\bf J=L+S}$, where $\bf L$ is
the orbital angular momentum, and ${\bf S}=\frac{1}{2}{\bf\Sigma}$
is the spin operator. The operator $K$ is defined as
$K=\gamma^0({\bf\Sigma}\cdot{\bf L}+1)$, which commutes with both
$H$ and {\bf J}. Explicitly, we have \be K&=&{\rm diag
}\left({\hat k},-{\hat k}\right),\nonumber\\ {\hat
k}&=&\bsigma\cdot {\bf L} +1. \ee
 The common eigenstates can be
written as \cite{Ho2}
 \be \psi=\frac{1}{r} \left(
\begin{array}{c} f_+(r) {\cal Y}^k_{jm_j}\\ if_-(r){\cal
Y}^{-k}_{jm_j}
\end{array}\right),
\ee here ${\cal Y}^k_{jm_j}(\theta,\phi)$ are the spin harmonics
satisfying
 \be
 {\bf J}^2 {\cal Y}^k_{jm_j} &=& j(j+1){\cal
Y}^k_{jm_j},~~j=\frac{1}{2},\frac{3}{2},\ldots ,
\\ J_z {\cal Y}^k_{jm_j} &=& m_j{\cal
Y}^k_{jm_j},~~~~~~~~|m_j|\leq j ,
\\ {\hat k}{\cal Y}^k_{jm_j}&=& -k{\cal Y}^k_{jm_j},~~~~~~~~~
k=\pm(j+\frac{1}{2}),
 \ee
 and
  \be ({\bsigma}\cdot {\hat {\bf
r}}){\cal Y}^k_{jm_j}= -{\cal Y}^{-k}_{jm_j}.
 \ee
 Using the identity (\ref{id}),
one gets
 \be {\bsigma}\cdot ({\bf p\pm i\mu {\bf
E}})=i({\bsigma}\cdot {\hat {\bf r}})\left(-\partial_r+ \frac
{{\hat k}-1}{r}\pm \mu E_r\right).
\ee
Eq.(\ref{H1}) then reduces
to
 \be
\left(\frac{d}{dr} + \frac{k}{r} + \mu E_r\right)f_+&=& \left(E +
M\right)f_- ,
\\ \left(-\frac{d}{dr} +
\frac{k}{r} + \mu E_r\right)f_- &=& \left( E - M\right)f_+ .
 \ee
These two equations have the same SUSY structure as (\ref{f+}) and
(\ref{f-}), and so one can proceed as in the last section. Again,
we consider the situation where $k<0$ and $\int dr \mu E_r
>0$, so that for unbroken supersymmetry, the ground state has
upper component $f_+^{(0)}$ normalizable, and lower component
$f_-^{(0)}=0$. The other situation can be discussed similarly. In
this case, eq.(\ref{w1}) becomes
 \be W^\prime=\frac{g}{r} -\mu E_, ~~~g\equiv |k|. \label{W2}
 \ee
Thus all the discussions following (\ref{w1}) can be carried over
with the change $A_\phi(r)\rightarrow \mu E_r(r)$.

Particularly, the case with $\mu E_r \propto r$ is just the Dirac
oscillator \cite{Dirac-osc}. Thus the new exactly solvable models
associated with the $X_\ell$ Laguerre polynomials constructed
according to the procedure in Sect.~III are the new deformed Dirac
oscillators.

\subsection{Dirac-Pauli equation with electric fields in
cylindrical coordinates}

We now turn to $2+1$ dimensions where the electric field is
cylindrically symmetric. This discussion can be readily extended
to the case in $3+1$ dimensions where the electric field is
constant along the $z$ direction.

The Dirac-Pauli Hamiltonian takes the form
 \be
H_{DP}={\bsigma}\cdot{\bf p}+i\mu \sigma_3{\bsigma}\cdot{\bf E}
+\sigma_3 M.\label{H-DP-2d}
 \ee
All vectors lie in the $x-y$ plane.  For ${\bf
E}=E_r(r)\hat{\bf{r}}$, the second term in (\ref{H-DP-2d}) is $ -
i\mu(\bsigma\cdot \hat{\bf{r}})\sigma_3 E_r(r) $ (note that
$\sigma_3\cdot \bsigma= - \bsigma \cdot \sigma_3$ in this case).
Comparing this term with $-\bsigma\cdot \bf{A}$ in (\ref{H-D}) and
(\ref{id-1}), one sees that the Hamiltonians (\ref{H-DP-2d}) and
(\ref{H-D}) are equivalent if we make the correspondence $\mu
E_r(r) \leftrightarrow A_\phi(r)$. Therefore, the results in
Sect.~III can be directly carried over to the present case.

We note here that the correspondence $\mu E_r(r) \leftrightarrow
qA_\phi(r)$ (we restore the charge $q$) also underlies the duality
between the Aharonov--Casher effect \cite{AC}, which is described
by the Dirac-Pauli equation, and the well-known Aharonov--Bohm
effect \cite{AB}, described by the Dirac equation with minimal
coupling. The Aharonov--Casher effect concerns the topological
phase experienced by a neutral fermion with a magnetic moment when
diffracted around a line of electric charge. It is an
electrodynamic and quantum--mechanical dual of the Aharonov--Bohm
effect, which gives a phase shift for a charged particle
diffracting around a tube of magnetic flux.

\subsection{Two-dimensional Dirac equation with Lorentz scalar
potential}

Let us consider a $(1+1)$-dimensional Dirac Hamiltonian
 of the kind
\begin{equation}
H=\alpha p + \beta (M+V_s(x))~ \label{H2}
\end{equation}
where $M$ is the mass of the fermion, $p=-id/dx$, $\alpha$ and
$\beta$ are the Dirac matrices,  and $V_s$ is the Lorentz scalar
potential. Such model is of interest in the theory of nuclear
shell model \cite{Bhad}, and as a model of the self-compatible
field of a quark system \cite{GMM}.

This system is supersymmetric \cite{SUSY1}, as can be easily shown
as follows. We represent the Dirac matrices by
\begin{eqnarray}
\alpha = \sigma_2,~~~ \beta = \sigma_1. \label{rep}
\end{eqnarray}
Then the Dirac equation $H\psi =E\psi$ for the two-component wave
function
\begin{eqnarray}
\psi(x)= \left( \begin{array}{c} \psi_+ (x)\\ \psi_- (x)
\end{array}\right)
\end{eqnarray}
takes the form
\begin{eqnarray}
\left(\frac{d}{dx} -W^\prime (x)\right)\psi_+ &=& E\psi_- ,\nonumber\\
\left(-\frac{d}{dx} -W^\prime (x)\right)\psi_- &=& E\psi_+ .
\label{susy-1a}
\end{eqnarray}
Here $W^\prime (x)\equiv -(V_s(x) + M)$ and $\mathcal{E}=E^2$.
Eq.~(\ref{susy-1a}) is now in the SUSY form, with $W(x)$ playing
the role of the prepotential.  As such, following the procedure in
Sect.~III, we can obtain new exactly solvable systems related to
the exceptional Laguerre polynomials.

In fact, in this case the domain need not be confined to the
half-line. Thus we can construct new solvable Dirac systems whose
eigenfunctions are related to the exceptional Jacobi polynomials
by simply linking $W$ with the prepotential corresponding to
exceptional Jacobi polynomials.

\section{Fokker-Planck equations}

Finally, we discuss briefly how the exceptional orthogonal
polynomials can appear in the Fokker-Planck (FP) equations.  In
one dimension, the FP equation of the probability density
$\mathcal{P}(x,t)$ is \cite{FP}
\begin{gather}
\frac{\partial}{\partial t} \mathcal{P}(x,t)=\mathcal{L}\mathcal{P}(x,t),\nonumber\\
\mathcal{L}\equiv -\frac{\partial}{\partial x} D^{(1)}(x) +
\frac{\partial^2}{\partial x^2}D^{(2)}(x). \label{FPE}
\end{gather}
The functions $D^{(1)}(x)$ and $D^{(2)}(x)$ in the FP operator
$\mathcal{L}$ are, respectively,  the drift and the diffusion
coefficient (we consider only time-independent case).  The drift
coefficient represents the external force acting on the particle,
while the diffusion coefficient accounts for the effect of
fluctuation.   Without loss of generality, in what follows we
shall take $D^{(2)}=1$.  The drift coefficient can be defined by a
prepotential $W(x)$ as $D^{(1)}(x)=2W^\prime (x)$.

The FP equation is closely related to the Schr\"odinger equation
\cite{HS,FP}.  Substituting
 \be
 \mathcal{P}(x,t)\equiv e^{-\lambda t} e^{W(x)} \phi(x).
 \ee
into the FP equation, we find that $\phi$ satisfies the
Schr\"odinger-like equation: $H\phi=\lambda\phi$, where
\begin{eqnarray}
H&\equiv& -e^{-W}\mathcal{L}e^{W} \nonumber\\
&=&-\frac{\partial^2}{\partial x^2}+W^\prime (x)^2 +
W^{\prime\prime}(x).\nonumber
\end{eqnarray}
Thus $\phi$ satisfies the time-independent Schr\"odinger equation
with Hamiltonian $H$ and eigenvalue $\lambda$, and
$\phi_0=\exp(W)$ is the zero mode of $H$: $H\phi_0=0$.

It is now clear that FP equations transformable to exactly
solvable Schr\"odinger equations can be exactly solved. If all the
eigenfunctions $\phi_n$ ($n=0,1,2,\ldots$) of $H$ with eigenvalues
$\lambda_n$ are solved, then the eigenfunctions $\mathcal{P}_n(x)$
of $\mathcal{L}$ corresponding to the eigenvalue $-\lambda_n$ is
$\mathcal{P}_n(x)=\phi_0(x)\phi_n(x)$.  The stationary
distribution is $\mathcal{P}_0=\phi_0^2=\exp(2W)$ (with $\int
\mathcal{P}_0(x)\,dx=1$), which is obviously non-negative, and is
the zero mode of $\mathcal{L}$: $\mathcal{L}\mathcal{P}_0=0$. Any
positive definite initial probability density $\mathcal{P}(x,0)$
can be expanded as $\mathcal{P}(x,0)=\phi_0(x)\sum_n
c_n\phi_n(x)$, with constant coefficients $c_n$ ($n=0,1,\ldots$)
\begin{eqnarray}
c_n=\int_{-\infty}^\infty \phi_n(x)\left(\phi_0^{-1}(x)
\mathcal{P}(x,0)\right)dx.
\end{eqnarray}
Then at any later time $t$, the solution of the FP equation is
$\mathcal{P}(x,t)=\phi_0(x)\sum_n c_n \phi_n(x)\exp(-\lambda_n
t)$.

Thus all the shape-invariant potentials in supersymmetric quantum
mechanics give the corresponding exactly solvable FP systems. One
needs only to link the prepotential $W$ in the Schr\"odinger
system with the drift potential corresponding to the drift
coefficient $D^{(1)}=2W^\prime$ in the FP system. For example, the
shifted oscillator potential in quantum mechanics corresponds to
the FP equation for the well-known Ornstein--Uhlenbeck process
\cite{FP}.

Of interest to us here is the FP equation that corresponds to the
radial oscillator potential.  This FP equation describes the
so-called Rayleigh process \cite{Rayleigh}.  When $W(x)$ is
replaced by the the prepotential (\ref{wl-osc}), we obtain an
exactly solvable FP equation, describing a deformed Rayleigh
process, whose eigenfunctions are given by the exceptional
$X_\ell$ Laguerre polynomials.

Again, as with the Dirac equation with Lorentz scalar potential,
in this case one can get exactly solvable FP equations whose
eigenfunctions are related to the exceptional Jacobi polynomials
by simply linking $W$ with the prepotential corresponding to
exceptional Jacobi polynomials.

\section{Summary}

The discovery of the exceptional $X_\ell$ Laguerre and Jacobi
polynomials has opened up new avenues in the area of mathematical
physics and in classical analysis. Unlike the well-known classical
orthogonal polynomials which start with constant terms, these new
polynomials have lowest degree $\ell=1,2,\ldots$, and yet they
form complete set with respect to some positive-definite measure.
Some properties of these new polynomials have been studied, and
many more have yet to be investigated.

In this paper we have presented some physical models in which
these new polynomials could play a role.  For clarity of
presentation, we concentrate mainly on the exceptional Laguerre
polynomials.  We show how some Dirac equations coupled minimally
and non-minimally with external fields, and the Fokker-Planck
equations can be exactly solvable with the exceptional Laguerre
polynomials as the main part of the eigenfunctions.  The systems
presented here have enlarged the number of exactly solvable
physical systems known so far.

\begin{acknowledgments}

This work is supported in part by the National Science Council
(NSC) of the Republic of China under Grant NSC
96-2112-M-032-007-MY3 and NSC-99-2112-M-032-002-MY3.  I thank S.
Odake and  R. Sasaki for many helpful discussions on the
exceptional polynomials.

\end{acknowledgments}

\end{document}